\documentstyle[aps,amssymb,multicol,epsfig]{revtex}

\begin{document}
\author{I.M. Sokolov$^{1}$, A. Blumen$^{1}$ and J. Klafter$^{2}$}
\address{$^{1}$Theoretische Polymerphysik, Universit\"{a}t Freiburg,\\
Hermann-Herder-Str.3, D-79104 Freiburg im Breisgau, Germany\\
$^{2}$School of Chemistry, Tel Aviv University, \\
69978 Tel Aviv, Israel}
\title{Linear Response in Complex Systems: CTRW and the Fractional Fokker-Planck
Equations. }
\date{\today}
\maketitle

\begin{abstract}
We consider the linear response of systems modelled by continuous-time
random walks (CTRW) and by fractional Fokker-Planck equations under the
influence of time-dependent external fields. We calculate the corresponding
response functions explicitely. The CTRW curve exhibits aging, {\it i.e.} it
is not translationally invariant in the time-domain. This is different from
what happens under fractional Fokker-Planck conditions.

\end{abstract}

\begin{multicols}{2}

\section{Introduction}

Many systems, such as polymer chains and networks, proteins, glasses and
charge-carriers in semiconductors are characterized by extremely slow
relaxation processes. The microscopic mechanisms leading to such slow
relaxations differ considerably from system to system, but their microscopic
manifestations often correspond to power-laws. A recently introduced
approach to slow relaxation in {\it time-independent} external fields is
based on fractional Fokker-Planck equations (FFPEs) \cite{MeKla,Metz1} or on
fractional Master equations \cite{AH}. The FFPE describing subdiffusive
behavior in an external field reads:

\begin{equation}
\frac{\partial }{\partial t}P(x,t)=\,_{0}D_{t}^{1-\gamma }{\cal L}%
_{FP}P(x,t),  \label{subdiff}
\end{equation}
where $P(x,t)$ is the pdf to find a particle (walker) at point $x$ at time $%
t $. In Eq.(\ref{subdiff}) $_{0}D_{t}^{1-\gamma }$ is the fractional
Riemann-Liouville operator ($0<\gamma <1$) defined through \cite{MeKla} 
\begin{equation}
_{0}D_{t}^{1-\gamma }Z(t)=\frac{1}{\Gamma (\gamma )}\frac{\partial }{%
\partial t}\int_{0}^{t}dt^{\prime }\frac{Z(t^{\prime })}{(t-t^{\prime
})^{1-\gamma }}.  \label{RiLi}
\end{equation}
Here $t=0$ can be associated with the time at which the system was prepared.
Furthermore, in Eq.(\ref{subdiff}) ${\cal L}_{FP}$ is the Fokker-Planck
operator, 
\begin{equation}
{\cal L}_{FP}=K\Delta -{\bf \mu }\nabla {\bf f}(x).  \label{FPOp}
\end{equation}
In what follows we concentrate on the case in which the acting force ${\bf f}%
(x)=-\nabla U$ is homogeneous and of magnitude ${\bf E}$. In this case we
have ${\cal L}_{FP}=K\Delta -{\bf \mu E}\nabla $.

The fractional Fokker-Planck equation, Eq.(\ref{subdiff}), has turned out to
be useful in describing a broad range of phenomena connected with anomalous
diffusion \cite{MeKla}. There are essentially two main mechanisms leading to
long-time memory in the behavior of complex systems. One is related to the
hierarchical structure of the modes of the system, as is the case for
polymer chains and networks \cite{Schiessel} and for rough interfaces \cite
{Liu}. The other is associated with the diffusion in complex, (almost)
quenched potential landscapes, which occur in glassy systems (ranging from
normal glasses to proteins). A phenomenological description of free
relaxation (i.e. the time-evolution of a system which is prepared in a
nonequilibrium initial condition and then evolves under time-independent
external conditions) in terms of fractional Fokker-Planck equations is
reasonable in both situations. Thus, the FFPE, Eq.(\ref{subdiff}) can be
viewed as a phenomenological linear-response theory for a system with long
memory in contact with a heat bath \cite{Sok1}. This equation can be derived
systematically from the continuous-time random walk (CTRW) scheme using the
standard Kramers-Moyal procedure. The equivalence of the two approaches was
discussed in \cite{BMK} using the subordination property of CTRWs; see also 
\cite{BaSi,Barkai}. Thus, the free relaxation properties of a CTRW system
are closely reproduced by Eq.(\ref{subdiff}) and can be expressed in terms
of Mittag-Leffler functions \cite{MeKla,AH}.

In what follows we concentrate on the response of the two models to {\it %
time-dependent} external fields, and discuss the mean current (velocity) and
mean polarization (coordinate) as a function of time in systems under pulsed
or sinusoidal external fields. We show that the linear response to
time-dependent fields predicted by FFPE (which is exactly the same as for
hierarchical models) differs strongly from what is expected under CTRW
conditions. Thus, the response to time-dependent fields of the systems
described by FFPE is mainly influenced by the values of the field at times
immediately preceding the observation time $t$, while the memory of the
earlier history fades out. On the other hand, in CTRW-systems the response
of the current to external fields is local in time, while the corresponding
susceptibility decays. In such systems the polarization depends much on the
early history of the system.

\section{Linear response within the FFPE scheme}

Let us consider a system whose dynamics is described by a fractional
Fokker-Planck equation and concentrate on the mean particles' displacement
under the action of an external force. This mean displacement $\overline{%
{\bf X}(t)}$ is given by 
\begin{equation}
\overline{{\bf X}(t)}=\int_{-\infty }^{\infty }{\bf x}P({\bf x},t)d{\bf x}.
\end{equation}
Multiplying Eq.(\ref{subdiff}) by $x$ and integrating it over the whole
space we obtain: 
\begin{eqnarray}
\frac{\partial }{\partial t}\int {\bf x}P({\bf x},t)d{\bf x}
&=&\,_{0}D_{t}^{1-\gamma }\left[ K\int {\bf x}\Delta P({\bf x},t)d{\bf x}%
\right.   \nonumber \\
&&-\left. {\bf \mu E}(t)\int {\bf x}\nabla P({\bf x},t)d{\bf x}\right] .
\label{part}
\end{eqnarray}
The left hand side of Eq.(\ref{part}) is nothing but $\frac{d}{dt}\overline{%
{\bf X}(t)}$, whereas the right hand side can be simplified by integration
by parts. Since $P({\bf x},t)$ and its derivatives with respect to the
coordinates vanish at infinity, the first integral vanishes and the second
one is unity. Hence 
\begin{equation}
\frac{d}{dt}\overline{{\bf X}(t)}=\,_{0}D_{t}^{1-\gamma }{\bf \mu E}(t).
\label{FracDer}
\end{equation}
Note that such kind of response is typical for complex and hierarchically
built systems, like polymer chains and networks, see \cite{Schiessel}.

As a simple example let us consider a chain of $N\gg 1$ beads connected by
harmonic springs and immersed in a viscous fluid (a Rouse-chain). We assume
the first bead of the chain to be tagged (say, charged) and experience the
external field ${\bf E}$. The motion of the beads is governed by the
equation 
\begin{equation}
\zeta \frac{d{\bf x}_{0}}{dt}=-k({\bf x}_{0}-{\bf x}_{1})+{\bf E}(t)+{\bf f}%
_{0}(t)  \label{Bead0}
\end{equation}
for the first bead (bearing number 0), 
\begin{equation}
\zeta \frac{d{\bf x}_{N}}{dt}=-k({\bf x}_{N}-{\bf x}_{N-1})+{\bf f}_{N}(t)
\label{Last}
\end{equation}
for the last one and 
\begin{equation}
\zeta \frac{d{\bf x}_{i}}{dt}=k({\bf x}_{i+1}+{\bf x}_{i-1}-2{\bf x}_{i})+%
{\bf f}_{i}(t)  \label{Bead1}
\end{equation}
for all other beads, $0<i<N$. In Eqs.(\ref{Bead0}) to (\ref{Bead1}) ${\bf f}%
_{i}(t)$ are Gaussian, $\delta $-correlated forces with zero mean. Note that
due to the linearity of Eqs.(\ref{Bead0}) to (\ref{Bead1}) the mean
positions and velocities of the beads (averaged over the realizations of the 
${\bf f}_{i}(t)$) follow equations similar in form, but where the random
forces ${\bf f}_{i}(t)$ are omitted. Let us suppose that the velocity
response of the first bead of the chain (averaged over the realizations of
the random forces ${\bf f}_{i}(t)$) is described by the memory function $%
M(t) $. Supposing that the initial mean velocities of the beads vanish, this
leads after a Laplace transformation to 
\begin{equation}
{\bf v}_{0}(\lambda )=\lambda {\bf x}_{0}(\lambda )=M(\lambda ){\bf E}%
(\lambda ).  \label{Lpl}
\end{equation}
If $N$ is very large, $N\rightarrow \infty $, the chain can be considered as
infinite; then the subchain starting with bead 1 has the same linear
response properties as the entire chain starting with bead 0. Thus, $M(t)$
can be found using Eq.(\ref{Bead0}) and noting that the equation of motion
for bead 1 has the same form, Eq.(\ref{Lpl}), with the acting force now
being ${\bf F}=k({\bf x}_{0}-{\bf x}_{1})$ instead of ${\bf E}$. The
Laplace-representation of the equation of motion for the first bead reads: 
\begin{equation}
\zeta {\bf v}_{0}(\lambda )=-k\left[ \frac{{\bf v}_{0}(\lambda )}{\lambda }-%
\frac{{\bf v}_{1}(\lambda )}{\lambda }\right] +{\bf E}(\lambda )  \label{L0}
\end{equation}
(where we set ${\bf x}_{0}(\lambda )={\bf v}_{0}(\lambda )/\lambda $ and $%
{\bf x}_{1}={\bf v}_{1}(\lambda )/\lambda $). Moreover, the analog of Eq.(%
\ref{Lpl}) for the velocity of the bead 1 reads: 
\begin{equation}
{\bf v}_{1}(\lambda )=M(\lambda )k\left[ \frac{{\bf v}_{0}(\lambda )}{%
\lambda }-\frac{{\bf v}_{1}(\lambda )}{\lambda }\right] .  \label{L1}
\end{equation}
The solution of Eqs.(\ref{Lpl}) to (\ref{L1}) gives $M(\lambda )=-\lambda
/2k+\sqrt{\lambda ^{2}/4k^{2}+\lambda /\zeta k}$. Thus, for $\lambda
\rightarrow 0$, $M(\lambda )\simeq \tau _{0}^{1/2}\sqrt{\lambda }$ with $%
\tau _{0}=1/\zeta k$, which is exactly the Laplace-representation of a
semi-derivative, $\tau _{0}^{1/2}\,_{0}D_{t}^{1-\gamma }$ with $\gamma =1/2$
in Eq.(\ref{FracDer}). This is the value which we will use in our numerical
examples in what follows.

In the case where the friction coefficients $\zeta _{i}$ and/or the spring
constants $k_{i}$ differ from site to site, other values of $\gamma $ may be
obtained \cite{Schiessel}. The same holds also for more complex
(higher-dimensional, fractal or tree-like) structures. For example, the case
when a tagged monomer is attached to a membrane corresponds to $\gamma =2/3$ 
\cite{Yoss}.

\section{Linear response of a CTRW system}

Let us now turn to the linear response in the framework of the CTRW model
introduced by Montroll and Weiss \cite{MoWei}. This model was extremely
successful in the explanation of dispersive transport in amorphous
semiconductors \cite{ScheM}; see \cite{HausKehr,BouGeor} for reviews. As is
usual in CTRW, we envisage an ensemble of noninteracting particles which may
be influenced by external fields (say, the particles are charged). The
particles follow then CTRWs, i.e. sequences of jumps. The time intervals $%
t_{i}$ between the jumps are uncorrelated. Of interest are waiting times
which follow power-law distributions, i.e. 
\begin{equation}
\psi (t)=\gamma /(1+t/\tau _{0})^{1+\gamma },\quad {\rm {with} \quad
0<\gamma <1.}  \label{wtd1}
\end{equation}
The physical motivation for such $\psi (t)$-forms may be rationalized using
random traps whose energy distribution is exponential \cite{BKZ}. In what
follows we put $\tau _{0}=1$ and work in dimensionless time units.

A basic quantity in the CTRW formalism is $\chi _{n}(t)$, the probability to
make exactly $n$ steps up to time $t$. In the standard, decoupled CTRW
picture (in which the spatial transition probabilities between the lattice
sites are independent of the waiting-times) under time-independent field the
probability distribution $P({\bf x},t)$ of finding a particle at ${\bf r}$
at time $t$ given that is started at ${\bf 0}$ at time 0 obeys \cite
{Blumofen} 
\begin{equation}
P({\bf x},t)=\sum_{i=0}^{\infty }P_{n}({\bf x})\chi _{n}(t),  \label{Green}
\end{equation}
where $P_{n}({\bf x})$ is the probability to reach ${\bf x}$ from ${\bf 0}$
in $n$ steps. We note that Eq.(\ref{Green}) states that the CTRW is a random
process subordinated to simple random walks under the operational time given
by the $\chi _{n}(t)$-distribution. Note that exactly this property is the
starting point for the derivation of FFPE in Ref. \cite{BMK}. Let us now
turn to the case of {\it time-dependent} fields. In this case the simple
subordination relation, Eq.(\ref{Green}) breaks down, since $P_{n}({\bf x})$
starts to depend on the actual value of external field at time instants of
steps, and thus on the actual times of steps, and not only on their number.
On the other hand, the mean velocity or mean displacement of the particles
can still be easily found.

Let us discuss the linear response of an ensemble of random walkers
performing CTRWs to a changing external field. We consider some physically
short time interval $dt$. Let $dN$ be the mean number of steps performed
during $dt$. The mean displacement during $dt$ is $\overline{d{\bf X}}={\bf x%
}dN$ where ${\bf x}$ is the mean displacement per step depending on the
actual value of the external field ${\bf E}$: 
\begin{equation}
{\bf x}={\bf \mu E}=\sum_{i}{\bf x}_{i}\frac{\left( {\bf x}_{i}\cdot {\bf E}%
\right) }{k_{B}T}{\bf .}  \label{Displ}
\end{equation}
Here ${\bf \mu }$ is the mobility tensor. The sum in the second expression
runs over all nearest neighbors vectors, $k_{B}$ is the Boltzmann constant
and $T$ is the temperature. We obtain now 
\begin{equation}
\overline{d{\bf X}}={\bf \mu E}(t)dN,  \label{Xmean}
\end{equation}
where $dN=N(t+dt)-N(t)\approx dt\sum_{i=0}^{\infty }n\frac{d}{dt}\chi
_{n}(t) $. Thus, the typical particles' velocity (which is proportional to
the particles' current) is given by: 
\begin{equation}
\overline{{\bf V}(t)}=\frac{\overline{d{\bf X}}}{dt}=f(t){\bf \mu E}(t).
\label{Ve}
\end{equation}
where 
\begin{equation}
f(t)=\sum_{i=0}^{\infty }n\frac{d}{dt}\chi _{n}(t).  \label{Nmean}
\end{equation}
Now the current density is ${\bf J}(t)=ne\overline{{\bf V}(t)}$, where $n$
is the density of charge carriers (assumed to be homogeneous) and $e$ is
their charge; the analogue of Eq.(\ref{Ve}) holds also for the current, 
\begin{equation}
{\bf J}(t)=f(t){\bf \sigma E}(t),  \label{Current}
\end{equation}
where ${\bf \sigma }=ne{\bf \mu }$ is the conductivity tensor.

According to the theory of CTRW, $\chi _{n}(\lambda )$, the
Laplace-transform of $\chi _{n}(t)$, reads $\chi _{n}(\lambda )=\psi
(\lambda )^{n}\left[ 1-\psi (\lambda )\right] /\lambda $, Ref.\cite{BKZ}.
From Eq.(\ref{Nmean}) we now have that the Laplace-transform of $f(t)$
reads: 
\begin{equation}
f(\lambda )=\frac{\psi (\lambda )}{1-\psi (\lambda )}  \label{Flapl}
\end{equation}
The Laplace transform of $\psi (t)$, Eq.(\ref{wtd1}), for small $\lambda $
is known to be $\psi (\lambda )=1-\lambda ^{\gamma }\Gamma (1-\gamma )$ \cite
{Shlesinger}. The inverse Laplace-transform of Eq.(\ref{Flapl}) (for longer
times) thus reads: 
\begin{equation}
f(t)=\frac{\sin \pi \gamma }{\pi }t^{\gamma -1}.  \label{N1}
\end{equation}
Note that the convergence to the behavior given by Eq.(\ref{N1}) can be very
fast. Moreover, effective interpolating forms valid both at short and at
longer times can be obtained. As an example let us consider the case when $%
\psi (t)$ is a one-sided (extreme) L\'{e}vy-law, $\psi (t)=L(t,\gamma
,-\gamma )$ (with $0<\gamma <1$), whose Laplace-transform is a stretched
exponential $\psi (u)=\exp (-\lambda ^{\gamma })$. This leads to $f(\lambda
)=\left[ \exp (\lambda ^{\gamma })-1\right] ^{-1}$. Note that the behavior
for large $\lambda $, i.e. short $t$, corresponds to that of $\psi (t)$,
since the denominator of $f(\lambda )$ is dominated by the exponential term.
For small $\lambda $ the asymptotic behavior given by Eq.(\ref{N1}) sets in.
The transition between the two types of behaviors takes place at $t\simeq 1$%
, and thus at longer times Eq.(\ref{N1}) holds.

Thus, for $t\gtrsim 1$ one has 
\begin{equation}
\overline{{\bf V}(t)}=\frac{\sin \pi \gamma }{\pi }t^{\gamma -1}{\bf \mu E}%
(t).  \label{Ve2}
\end{equation}
A similar equation holds, of course, for the current ${\bf J}(t)$ flowing
through the system: 
\begin{equation}
{\bf J}(t)=\frac{\sin \pi \gamma }{\pi }t^{\gamma -1}{\bf \sigma E}(t).
\label{CTRWCurr}
\end{equation}
The current response of a CTRW-system to an external field is local in time
and depends explicitly on the time elapsed after the system was prepared.
Systems in which the response to an external agent depends explicitly on the
delay between preparation time and measurement time are referred to as aging
systems. This kind of behavior is found to be pronounced in CTRWs with $%
0<\gamma <1$ \cite{Feigelman,Bouchaud,Montus,Maas}.

To obtain the particle's position and therefore the polarization of the
medium we simply have to integrate Eq.(\ref{Ve2}) over time. Hence: 
\begin{equation}
\overline{{\bf X}(t)}=\frac{\sin \pi \gamma }{\pi }\int_{0}^{t}t_{1}^{\gamma
-1}{\bf \mu E}(t_{1})dt_{1}
\end{equation}
or 
\begin{equation}
\overline{{\bf P}(t)}=\frac{\sin \pi \gamma }{\pi }\int_{0}^{t}t_{1}^{\gamma
-1}{\bf \sigma E}(t_{1})dt_{1}.  \label{CTRWRes}
\end{equation}
In the limit $t\rightarrow \infty $ these expressions for $\overline{{\bf X}%
(t)}$ or $\overline{{\bf P}(t)}$ are Mellin-transforms of the external
field, i.e.: 
\begin{equation}
{\bf P}_{\infty }={\bf \sigma }\frac{\sin \pi \gamma }{\pi }{\cal M}[{\bf E}%
;\gamma ].
\end{equation}
where ${\cal M}[f;s]=\int_{0}^{\infty }f(t)t^{s-1}dt$. Thus, the response of
the CTRW-system to a time-dependent field decays, and its polarization tends
to a constant value. Interestingly, the CTRW-system not only ages, but shows
a kind of ''Freudistic'' response: the polarization at time $t$ is mainly
due to the early history of the system, immediately after it was prepared in
its initial state.

Let us compare the linear response (polarization vs. external field) of a
CTRW-system and of a system described by the FFPE. The current through a
system described by the FFPE is given by a fractional derivative, 
\begin{equation}
{\bf J}(t)=\frac{{\bf \sigma }^{*}}{\Gamma (\gamma )}\frac{d}{dt}%
\int_{0}^{t}(t-t_{2})^{\gamma -1}{\bf E}(t_{2})dt_{2}  \label{CurFFPE}
\end{equation}
and the polarization of the system by a fractional integral 
\begin{equation}
{\bf P}(t)=\frac{{\bf \sigma }^{*}}{\Gamma (\gamma )}\int_{0}^{t}(t-t_{2})^{%
\gamma -1}{\bf E}(t_{2})dt_{2}.  \label{PolFFPE}
\end{equation}
For the constant field ${\bf E}(t_{1})={\bf E}_{0}$ both expressions, Eqs. (%
\ref{CTRWRes}) and (\ref{PolFFPE}), describe essentially the same
time-evolution (if one sets ${\bf (\sigma }\sin \pi \gamma )/\pi ={\bf %
\sigma }^{*}/\Gamma (\gamma )=\sigma _{0}$), since Eqs. (\ref{CTRWRes}) and (%
\ref{PolFFPE}) are equivalent to each other under the change of variable $%
t_{2}=t-t_{1}$. This equivalence doesn't hold anymore for the
time-dependent-field. Note that the main difference of the response, Eq.(\ref
{PolFFPE}) as compared with Eq.(\ref{CTRWRes}), is the fact that here the
polarization is affected mainly by the latest events, and that the memory of
the early history of the system fades away.

\section{Examples}

As examples we discuss two simple situations of the response of CTRW and of
FFPE systems to external fields. We first choose a rectangular pulse
switched on at $t=t_{w}$ and off at $t=t_{z}$ and then a sinusoidal field
switched on at $t=0.$ We consider here a highly symmetric system described
by a scalar conductivity $\sigma $, in which the current flows in the
direction of the external field.

For CTRW Eq.(\ref{CTRWCurr}) gives

\begin{equation}
J(t)=\left\{ 
\begin{array}{ll}
0 & t<t_{w} \\ 
\sigma _{0}Et^{\gamma -1} & t_{w}\leq t\leq t_{z} \\ 
0 & t>t_{z}
\end{array}
\right.  \label{agedresp}
\end{equation}
which is a causal response concentrated on the time interval in which the
field acts. On the other hand, from Eq.(\ref{CurFFPE}) it follows for FFPE
that 
\begin{equation}
J(t)=\left\{ 
\begin{array}{ll}
0 & t<t_{w} \\ 
\sigma _{0}E(t-t_{w})^{\gamma -1} & t_{w}\leq t\leq t_{z} \\ 
\sigma _{0}E\left[ \left( t-t_{w}\right) ^{\gamma -1}-\left( t-t_{z}\right)
^{\gamma -1}\right] & t>t_{z}
\end{array}
.\right.  \label{Fracresp}
\end{equation}

Equations (\ref{agedresp}) and (\ref{Fracresp}) coincide only if one takes $%
t_{w}=0$, $t_{z}\rightarrow \infty $. This limit parallels the findings of
Ref. \cite{BMK}. In general, however, they are different. Eq.(\ref{agedresp}%
) describes a response concentrated on the time-interval of the
field-action, $t_{w}\leq t\leq t_{z}$: no afteraction effects are seen. The
current never changes sign and has finite jumps at $t=t_{w}$ and $t=t_{z}$.
On the other hand, Eq.(\ref{Fracresp}) shows considerable afteraction: the
current does not vanish for $t>t_{z}$. The current diverges at $t=t_{w}$ and 
$t=t_{z}$ and changes its sign from positive to negative at $t=t_{z}$.
Furthermore, the overall response described by Eq.(\ref{Fracresp}) is
invariant under time translation, i.e. it depends only on the differences $%
t-t_{w}$ and $t-t_{z}$, which is not the case for CTRW, Eq.(\ref{agedresp}).

\narrowtext
\begin{figure}[htb]
\epsfxsize=8truecm
\epsfbox{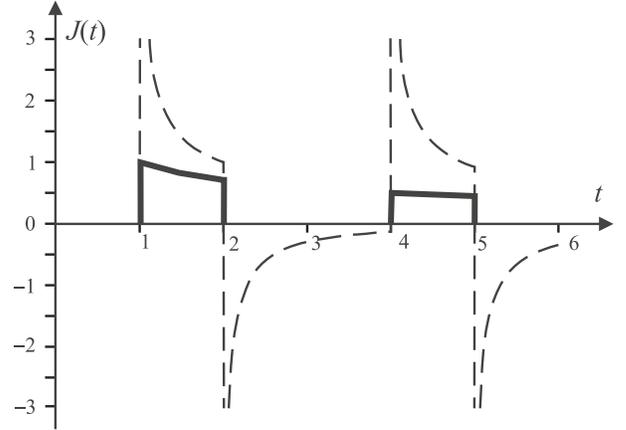}
\caption{ Shown is the current $J(t)$ in response to two
rectangular field pulses of unit amplitude acting during the intervals $%
1<t<2 $ and $4<t<5$. The thick solid line reprersents the CTRW-response,
while the dashed line represents the response of a system described FFPE,
see text for details.}
\label{fig1}
\end{figure}

Fig.1 shows the systems' response to the two pulses of unit length and unit
amplitude following at times $t_{1}=1$ and $t_{2}=4$ as follows from Eq.(\ref
{CTRWCurr}) and Eq.(\ref{CurFFPE}). The parameters are $\sigma _{0}=1$ and $%
\gamma =1/2$. The response to the first pulse follows exactly Eqs.(\ref
{agedresp}) and (\ref{Fracresp}); hence the differences between CTRW and
FFPE can be seen clearly. The comparison of the response of the two systems
to a second pulse is also very instructive. For systems described by the
FFPE the response to the first and to the second pulses do not differ (note
that the superposition principle holds and that the responses are additive).
On the other hand, in the case of a CTRW-system the second pulse causes a
much weaker reaction than the first one. The susceptibility of the system
(the conductivity) decays with time.

Let us now turn to the response of the system to a sinusoidal force switched
on at $t=0$, $E(t)=E_{0}\sin (\omega t)\theta (t)$. The current through the
quasiequilibrium (FFPE) system is described by the corresponding fractional
derivative, $J(t)\propto \,_{0}D_{t}^{1-\gamma }E(t)$. For example, for $%
\gamma =1/2$, the response to the sinusoidal field is given by: 
\begin{eqnarray}
J(t) &\propto &\frac{d^{1/2}}{dt^{1/2}}E(t)\propto \sigma _{0}E_{0}\omega
^{1/2}\left[ \sin (\omega t+\pi /4)\right.   \nonumber \\
&&-\left. \sqrt{2}{\rm Gres}(\sqrt{\omega t})\right] ,  \label{Eq.(29)}
\end{eqnarray}
whereas 
\begin{eqnarray}
P(t) &\propto &\frac{d^{-1/2}}{dt^{-1/2}}E(t)\propto \sigma _{0}E_{0}\omega
^{-1/2}\left[ \sin (\omega t-\pi /4)\right.   \nonumber \\
&&+\left. \sqrt{2}{\rm Fres}(\sqrt{\omega t})\right] .  \label{Eqn.30}
\end{eqnarray}
In Eqs.(\ref{Eq.(29)}) and (\ref{Eqn.30}) ${\rm Gres}(x)$ and ${\rm Fres}(x)$
are the auxiliary Fresnel integrals; see Ref. \cite{Oldham}. Since both
functions vanish for $x$ large, the response of the system ($J$ or $P$) at
long times tends to be sinusoidal, with a phase shift of $\pi /4$ or $-\pi /4
$ with respect to the acting force. The corresponding functions are shown in
the upper panel of Fig.2 for $\sigma _{0}=E_{0}=\omega =1$.

\narrowtext
\begin{figure}[htb]
\epsfxsize=8truecm
\epsfbox{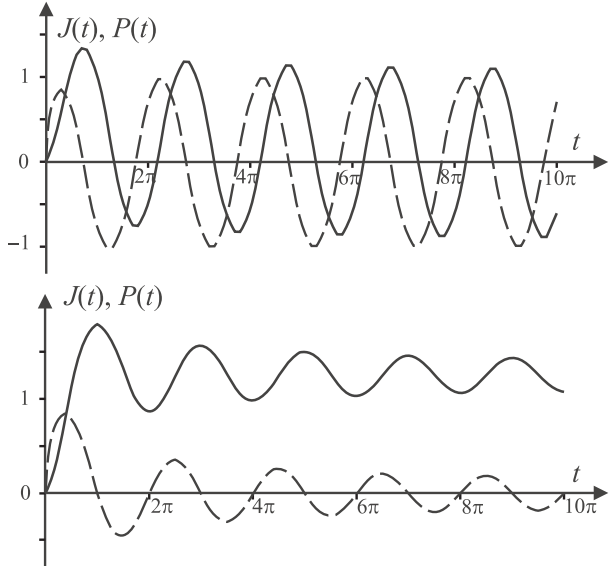}
\caption{The current $J(t)$ (dashed lines) and the
polarization $P(t)$ (full lines) as a response to the sinusoidal field $%
E(t)=\sin t$ for a system described by FFPE (upper panel) and by the
CTRW-model (lower panel), see text for details.}
\label{fig2}
\end{figure}

For a CTRW system under the same sinusoidal external field the current
through the system decays as 
\begin{equation}
J(t)\propto \sigma _{0}t^{-1/2}E(t)\propto \sigma _{0}E_{0}t^{-1/2}\sin
\omega t,  \label{Eq.(31)}
\end{equation}
while the polarization follows: 
\begin{eqnarray}
P(t) &=&\int_{0}^{t}J(t)dt\propto \sigma _{0}E_{0}\int_{0}^{t}t^{-1/2}\sin
\omega tdt  \label{Eq.(32)} \\
&=&\sigma _{0}E_{0}\sqrt{\frac{2\pi }{\omega }}S(\sqrt{\omega t}),  \nonumber
\end{eqnarray}
where $S(x)$ is the Fresnel integral. Note that since $\lim_{x\to \infty
}S(x)=1/2$, as time grows the polarization tends to a constant value $%
P_{\infty }\propto E_{0}/\sqrt{\omega }$ as time grows. The behavior of $J(t)
$ and $P(t)$, Eqs.(\ref{Eq.(31)}) and (\ref{Eq.(32)}) is shown in the lower
panel of Fig.2 for the same values of parameters as before.

\section{Conclusions}

We have considered the linear response of systems governed by CTRW and by
FFPE dynamics to time-dependent external fields. The form of the response
for cases described by CTRW displays a ''Freudistic'' memory and no
afteraction after switching off the field. This differs considerably from
the responce shown by systems obeying FFPE.

\section{Acknowledgments}

The authors gratefully acknowledge financial support by the GIF, by the
Deutsche Forschungsgemeinschaft and by the Fonds der Chemischen
Industrie.

\end{multicols}

\end{document}